\documentclass[pra,11pt]{revtex4-1}

\usepackage{amsmath}    
\usepackage{graphicx}   
\usepackage{verbatim}   
\usepackage{color}      
\usepackage{epsfig}
\usepackage{subfigure}  
\usepackage{hyperref}   

\newcommand{\bea}{\begin{eqnarray}}
\newcommand{\eea}{\end{eqnarray}}
\newcommand{\ket}[1]{\left|{#1}\right\rangle}
\newcommand{\bra}[1]{\left\langle{#1}\right|}
\newcommand{\aver}[1]{\left\langle{#1}\right\rangle}

\newcommand{\modu}[1]{\left|{#1}\right|}

\newcommand{\trev}{T_{\rm rev}}

\begin{document}
\title{Visualizing revivals and fractional revivals in a Kerr medium using optical tomogram}
\author{M. Rohith and C. Sudheesh}
\affiliation{Department of Physics, Indian Institute of Space Science and Technology, Thiruvananthapuram, 695 547, India.}



\begin{abstract}
We theoretically study the optical tomography of the time evolved states generated by the evolution of different kinds of  initial wave packets in a Kerr medium. Exact analytical expression for the optical tomogram of the quantum state at any instant during the evolution of a generic initial wave packet is derived in terms of Hermite polynomials. Time evolution of the optical tomogram is discussed for three kinds of  initial states:  a coherent state, an $m$-photon-added coherent state, and even and odd coherent states. We  show the   manifestation of revival and fractional revivals in the optical tomograms of the time evolved states. We find that the optical tomogram of the time evolved state at the instants of fractional revivals  shows structures with sinusoidal strands. The number of sinusoidal strands in the optical tomogram of the time evolved state at $l$-sub-packet fractional revivals is  $l$ times the number of sinusoidal strands present in the optical tomogram of the initial state.
We have also investigated the effect of decoherence on the optical tomograms of the states at the instants of fractional revivals for the initial states considered above. We consider amplitude decay and phase damping models of decoherence, and  show the direct manifestations of decoherence in  the optical tomogram.
\end{abstract}

\maketitle
\section{Introduction}
Time evolution of an initial wave packet in a nonlinear medium can exhibit revival and fractional revivals at specific instants of time. The revival phenomenon has been investigated both theoretically and experimentally in a wide class of systems \cite{robi}. A \textit{revival} of a well localized initial wave packet occurs when it evolves in time to a wave packet that reproduces the initial wave form. The characteristic time scale over which this phenomenon happens is called  the revival time $\trev$. Within this characteristic time scale $\trev$, the wave packet may split into number of scaled copies of initial state at specific instants during the evolution. This is known as the \textit{fractional revival} of the initial wave packets \cite{aver}. An $l$-sub-packet fractional revival occurs when the initial wave packet splits into superposition of $l$ wave packets of the initial form. The revivals and fractional revival has been observed experimentally in a variety of quantum systems such as Rydberg atomic wave packets, molecular vibrational states, Bose-Einstein condensates, and so forth \cite{rempe, yeazell1990, yeazell, Meacher, vrakking, greiner, Matsukevich}. Fractional revivals occurring in a nonlinear media can be used to generate various kinds of macroscopic superposition states of light. Such superposed states of light have potential applications in quantum optics and quantum information. Generation of discrete superposition of coherent states at fractional revival times in the process of wave packet propagation in a nonlinear media have been discussed in \cite{yurke,miranowicz,paprzycka,tara}. It has been shown that the superposed wave packets generated at fractional revival times, using an initial coherent state in a nonlinear medium, have application in quantum cloning \cite{cerf}. Two and four superposition states generated at fractional revival instances are useful for implementing the one- and two-bit logic gates \cite{shapiro}. Recent experimental observation of multicomponent Schr\"{o}dinger cat states using single-photon Kerr effect, opens up new directions for continuous variable quantum computation \cite{kirchmair}.

The signatures of fractional revivals in the time evolution of various physical quantities have been investigated theoretically \cite{jex, vaccaro, miranowicz2, sudheesh2004,romera2007,romera2008,rohith2014}.  The experimental characterization of time evolved states in a nonlinear media is an important aspect in the study of revival and fractional revivals and it  can be done by optical tomography, which is an efficient technique to measure and reconstruct the quantum state of optical fields \cite{leonhardt}. Optical tomography is based on one-to-one correspondence between the quasiprobability distribution and probability distribution of rotated quadrature phases of the field \cite{vogel}. The optical tomogram contains all the information about the system, and can serve as an alternative representation of the quantum system, apart from the conventional state vector or its density matrix representation in appropriate Hilbert space.
Optical tomogram of a quantum state can be theoretically calculated using  suitable transformations in symplectic tomogram \cite{mancini1995, ariano1996,omanko1997,mancini1996,manko1999,manko2005,manko2006} of  the quantum state. 
 In fact, a new formulation of quantum mechanics in which the quantum states are described by tomographic probability distributions  was suggested in \cite{Ibort}.  In experiments, a series of homodyne measurements of the rotated quadrature operator of the field are done on an ensemble of identically prepared systems. The quadrature histogram obtained by this method is called an optical tomogram. 
The first experimental observation of squeezed state of light, by measuring the quadrature amplitude distribution using the balanced homodyne detection arrangement, has been done in \cite{smithey}.  Thereafter, many nonclassical states of light have been characterized by optical homodyne tomography. A  review of continuous-variable optical quantum state tomography, including a list of the optical quantum states characterized by the same, is given in \cite{lvovsky}. 

It is a usual practice in experiments to reconstruct the density matrix or the quasiprobability distributions of the system from the optical tomogram and study its nonclassical properties. The reconstructed quasiprobability distributions like, Wigner function, Husimi $Q$-function, etc, provides a convenient way to visualize the fractional revivals in phase space. Recently, the quantum state collapse and revival due to single-photon Kerr effect has been observed using a three-dimensional circuit quantum electrodynamic architecture,  and  the multicomponent Sch\"{o}dinger cat states generated at fractional revival times are visualized in phase space using the quasiprobability distributions reconstructed from the optical tomogram \cite{kirchmair}. It should be emphasized that no reconstruction process  is perfect and the original errors of the experimental data can grow during the process of reconstruction. The physical properties of quantum states can be studied directly using optical tomogram  and  the tomographic approach can
be used to estimate the  errors in the histograms of experimentally obtained quadrature values \cite{bellini}. 
The macroscopic superposition states generated at the instants of fractional revivals are sensitive to interaction with its environment in an actual experimental settings, and this interaction can even destroy the states generated.
Aim of this paper is two fold. Firstly, to find the signatures of revivals and fractional revivals directly in the optical tomogram, which in turn can help experimentalists to  avoid the errors that can accumulate during the reconstruction process. Secondly,  to study the effects of amplitude decay and phase damping models of decoherence on the optical tomogram of the states at the instants of fractional revivals.
For this purpose, we consider  a nonlinear medium, which  models the wave packet propagation in a Kerr-like media \cite{milburn,kita} and the dynamics of Bose-Einstein condensates \cite{greiner}.   This paper is organized as follows: In section \ref{representation}, we give a brief review of  the tomographic representation of a quantum system. In section \ref{tomographic}, we theoretically calculate the optical tomogram of the time evolved states in the nonlinear medium. Here we discuss the evolution of optical tomogram for three specific initial states, which are, a coherent state,  an $m$-photon-added coherent state, and even and odd coherent states. Section \ref{decoherence} describes the effect of amplitude loss and phase noise on the optical tomograms of the  states at the instants of fractional revivals. In section \ref{conclusion}, we conclude the main results of this paper.

\section{Tomographic representation of quantum state}
\label{representation}

Optical tomography of several nonclassical states of light have been theoretically investigated in the literature \cite{filippov,korennoy,adam,rohith3}. A brief discussion about the calculation of optical tomogram of a  quantum state, and the  general properties of optical tomogram  are given below.  Consider the homodyne quadrature operator 
\begin{equation}
\hat{X}_{\theta}= \frac{1}{\sqrt{2}}\left(a\, e^{-i\theta}+a^\dag  e^{i\theta}\right),
\end{equation}
where $\theta$ is the phase of local oscillator in homodyne detection setup, and  $a$ and $a^\dag$ are the photon annihilation and creation operators of the single mode electromagnetic field, respectively. The phase of the local oscillator varies in the domain $0\leq \theta \leq 2\pi$. 
The optical tomogram $\omega\left(X_{\theta},\theta\right)$ of a quantum state with density matrix $\rho$  can be calculated by the following expression \cite{vogel,lvovsky}:
\begin{eqnarray}
\omega\left(X_{\theta},\theta\right)=\bra{X_{\theta},\theta}\rho\ket{X_{\theta},\theta},
\label{opt_tomo_def}
\end{eqnarray}
where
\bea
\ket{X_{\theta},\theta}=\frac{1}{\pi^{1/4}} \exp\left[-\frac{{X_{\theta}}^2}{2}-\frac{1}{2} e^{i\,2\theta} {a^\dag}^2+\sqrt{2}\, e^{i\,\theta} X_{\theta}\, a^\dag\right]\ket{0}\nonumber
\eea
is the eigenvector of the Hermitian operator $\hat{X}_{\theta}$ with eigenvalue $X_{\theta}$ \cite{barnett}.
 For a pure quantum state with wave vector $\ket{\psi}$, the expression in Eq.~(\ref{opt_tomo_def}) can be rewritten as
\begin{eqnarray}
\omega(X_{\theta},\theta)=\modu{\bra{X_{\theta},\theta}\psi\rangle}^2.\label{opt_tomo_def_purestate}
\end{eqnarray}
Normalization condition of the optical tomogram $\omega (X_{\theta},\theta)$ is given by
\bea
\int dX_\theta\, \omega (X_{\theta},\theta)=1.
\eea
The optical tomogram $\omega (X_{\theta},\theta)$ of a quantum state is non-negative and has the following symmetry property:  
\bea
\omega (X_{\theta},\theta+\pi)=\omega (-X_{\theta},\theta).
\eea
In subsequent sections, we use the Eq.~(\ref{opt_tomo_def}) to evaluate the optical tomogram of the quantum states
generated by Kerr medium.

\section{Optical tomography in a nonlinear medium}
\label{tomographic}
Consider the dynamics of a single-mode field governed by a nonlinear Hamiltonian 
\begin{equation}
H=\hbar \chi a^{\dag ^2}{a}^2=\hbar \chi N(N-1),
\label{kerrhamiltonian}
\end{equation}
where $a$ and $a^\dag$ are the photon annihilation and creation operators, respectively. The eigenstates of the operator $N=a^\dag a$  are the Fock state ${\ket{n}}$, where $n=0,\,1,\,2,...\,\infty$. The positive constant $\chi$ merely sets the time scale in the problem. We choose the numerical value of $\chi$ to be $5$ throughout this paper. 
Consider  a general initial wave packet $\ket{\psi(0)}$, with its Fock state expansion
\begin{equation}
\ket{\psi(0)}=\sum_{n=0}^{\infty} C_n \ket{n}\label{initial},
\end{equation}
where $C_n$ are the Fock state expansion coefficients. The time evolution of the state is governed by the Schr\"{o}dinger equation
\begin{equation}
\ket{\psi(t)}=U(t)\ket{\psi(0)} \label{Genpsi(t)},
\end{equation}
where $U(t)=\exp\left[-i H t/\hbar\right]$ is the unitary time evolution operator. The time evolved state at time $t$ can be written as 
\begin{equation}
\ket{\psi(t)}=\sum_{n=0}^{\infty} C_n e^{-i\chi t n (n-1)} \ket{n}.\label{general_psi(t)}
\end{equation}   
We theoretically calculate the optical tomogram of the time evolved state $\ket{\psi(t)}$, and look for the signatures of revival and fractional revival in the optical tomogram. Inserting Eq.~(\ref{general_psi(t)}) in Eq.~(\ref{opt_tomo_def_purestate}), we get the optical tomogram of the time evolved state $\ket{\psi(t)}$ as
\begin{eqnarray}
\omega\left(X_{\theta},\theta,t\right)=\frac{e^{-X_{\theta}^2}}{\sqrt{\pi}}\modu{\sum_{n=0}^{\infty}\frac{C_n\,e^{-i\chi t n(n-1)}}{\sqrt{n!} \,2^{n/2}}e^{-in\theta} H_n\left(X_{\theta}\right)}^2.\label{Genopticaltomo}
\end{eqnarray}
where $H_n(\cdot)$ denotes the Hermite polynomial of order $n$. Equation (\ref{Genopticaltomo}) gives the time evolution of the optical tomogram for an initial wave packet $\ket{\psi(0)}$ in a nonlinear medium modelled by the Hamiltonian $H$. In subsequent subsections, we discuss the temporal evolution of the optical tomogram for three different kinds of initial states, namely, a coherent state, an $m$-photon-added coherent states, and even and odd coherent states. 

\subsection{Evolution of coherent state}
\begin{figure}
\centering
\includegraphics[height=8 cm, width= 12 cm]{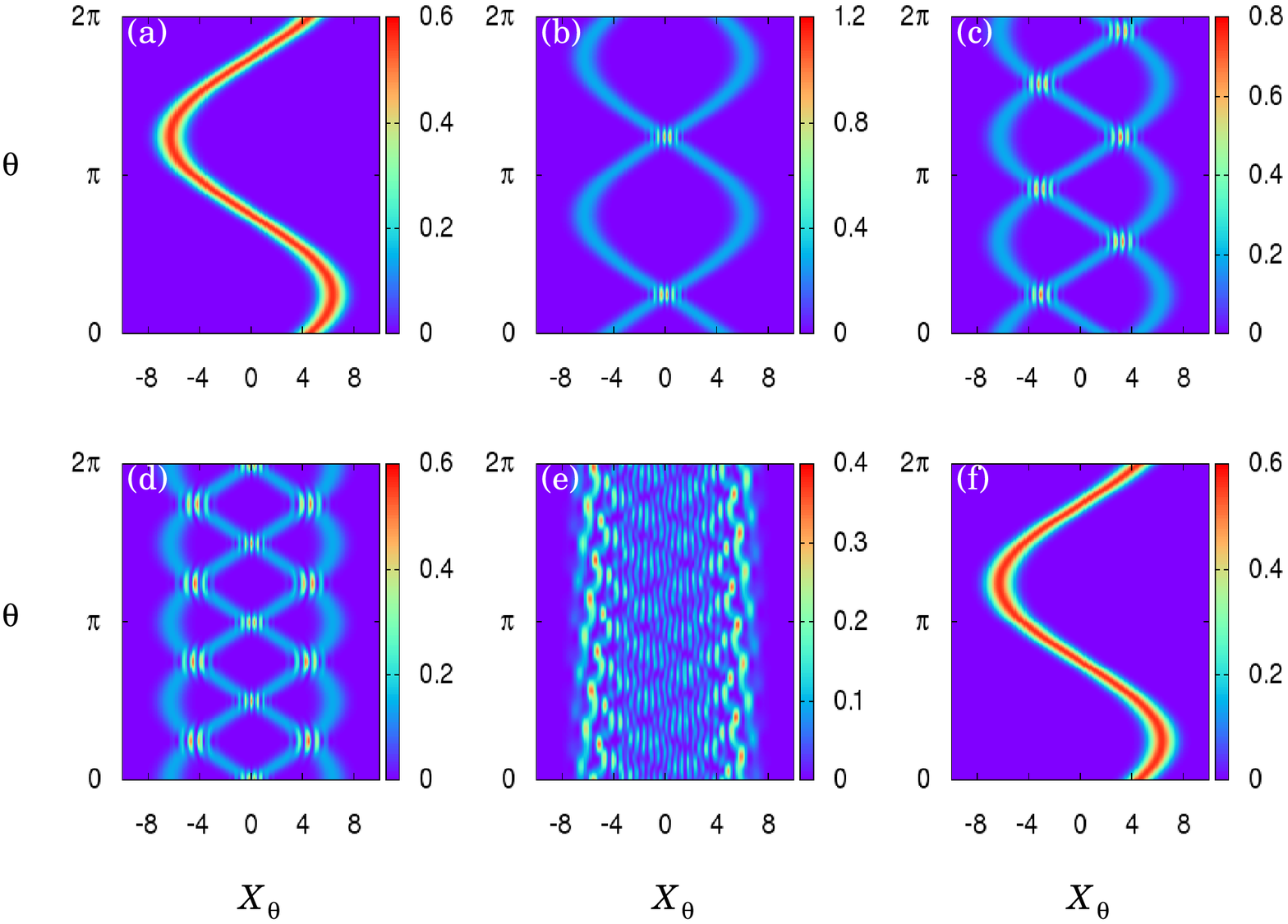} 
\caption{Time evolved optical tomogram $\omega_{\alpha}\left(X_{\theta},\theta,t\right)$ for an initial coherent state $\ket{\alpha}$ at  (a) $t=0$, (b) $t=\trev/2$, (c) $t=\trev/3$, (d) $t=\trev/4$, (e) $t=\trev/\sqrt{2}$ and (f) $\trev$, with field strength $\modu{\alpha}^2=20$. At $l$-sub-packet fractional revival time $t=\pi/l\chi$, the optical tomogram of the state shows structures with $l$ sinusoidal strands. The structures with sinusoidal strands are completely absent in the optical tomogram for the collapsed state at time $t=\trev/\sqrt{2}$.}
\label{fig:optCS}
\end{figure}

Consider the evolution of an initial coherent state $\ket{\alpha}$ in the nonlinear medium governed by the Hamiltonian in Eq.~(\ref{kerrhamiltonian}). The Fock state expansion coefficient $C_n$ in Eq.~(\ref{initial}) for the coherent state is 
\begin{equation}
C_n=e^{-\modu{\alpha}^2}\frac{\alpha^n}{\sqrt{n!}} .\label{CS}
\end{equation}
Let $\alpha=\sqrt{\modu{\alpha}^2}\,\exp(i\delta)$, where $|\alpha|^2$ is the mean number of photons in the coherent state $\ket{\alpha}$ and $\delta$ is a real number. Without loosing generality, we fix $\delta=\pi/4$. Figure \ref{fig:optCS}(a) displays the optical tomogram of the coherent state $\ket{\alpha}$ (at time $t=0$), for which the optical tomogram is given by
\begin{eqnarray}
\omega_{\alpha}\left(X_{\theta},\theta,t=0\right)=\frac{1}{\sqrt{\pi}} \exp\left[-\left(X_{\theta}-\sqrt{2}\modu{\alpha}\cos(\delta-\theta)\right)^2\right].
\end{eqnarray}
The maximum intensity of this optical tomogram $\omega_{\alpha}\left(X_{\theta},\theta,t=0\right)$ is $1/\sqrt{\pi}$, which occurs along the sinusoidal path, defined by $X_{\theta}=\sqrt{2\modu{\alpha}^2}\cos(\theta-\delta)$, in the $X_{\theta}-\theta$ plane. Hence, the projection of the optical tomogram on to $X_{\theta}-\theta$ plane is a structure with single sinusoidal strand. Along the $X_{\theta}$-axis ($\theta=0$), the maximum intensity of the optical tomogram occurs at $X_{\theta}=\sqrt{2\modu{\alpha}^2}\cos\delta$. 

The optical tomogram of the state at any instant during the evolution of a coherent state $\ket{\alpha}$ is calculated using Eq.~(\ref{Genopticaltomo}):
\begin{eqnarray}
\omega_{\alpha}\left(X_{\theta},\theta,t\right)=\frac{e^{-\modu{\alpha}^2}\,e^{-X_{\theta}^2}}{\sqrt{\pi}}
\modu{\sum_{n=0}^{\infty}\frac{\alpha^n\,e^{-i\chi
t n(n-1)}}{n!\,2^{n/2}}e^{-in\theta}
H_n\left(X_{\theta}\right)}^2.
\label{optCS}
\end{eqnarray}
Now, we analyse the optical tomogram $\omega_{\alpha}\left(X_{\theta},\theta,t\right)$ at the instants of fractional revivals.  In between $t=0$ and $t=T_{\rm rev}$, for an initial coherent state, $l$-sub-packet fractional revivals occur at time $t=\pi j /l \chi$, where  $j=1,2,\dots,(l-1)$ for a given value of   $l(>1)$  with a condition that $j$ and $l$ are mutually prime integers. 
Without loss of generality, we take $j=1$. The analysis shown below is true for any possible value of $j$.  The interesting periodicity properties of the unitary time evolution operator $U$ in Eq.~(\ref{Genpsi(t)}) at the instants of $l$-sub-packet fractional revivals, that is, at times $t=\pi/l\chi$, enables us to write the time evolved state $\ket{\psi(t)}$ at these instants as  
\begin{eqnarray}
\begin{split}
\ket{\psi_l}&=\ket{\psi(\pi/l\chi)} \\
&= \left\{ 
  \begin{array}{l l}
    \sum_{r=0}^{l-1} f_{r}^{(o)}\,\ket{\alpha\, e^{-i\,2 \pi r /l }} & \quad \text{if $l$ is odd}\\
    \sum_{r=0}^{l-1} f_{r}^{(e)}\, \ket{\alpha\,e^{-i\,\pi (2 r -1)/l }} & \quad \text{if $l$ is even,}
  \end{array} \right.
\end{split}  
  \label{cs_l}
\end{eqnarray}
where 
\bea
f_{r}^{(o)}&=&\frac{1}{l}\sum_{k=0}^{l-1}\exp\left[\frac{2\pi i r}{l}k\right]\exp\left[-\frac{i\pi}{l}k(k-1)\right], {\rm and}\\
f_{r}^{(e)}&=&\frac{1}{l}\sum_{k=0}^{l-1}\exp\left[\frac{2\pi i r}{l}k\right]\exp\left[-\frac{i\pi}{l}k^2\right]
\eea
are the Fourier coefficients \cite{tara}.  Note that, each of the ket vectors in Eq.~(\ref{cs_l}) is a coherent state and hence, the state $\ket{\psi_l}$ is a superposition of $l$ coherent states. At $l$-sub-packet fractional revival time $t=\pi/l\chi$, the Eq.~(\ref{optCS}) can be simplified to get the optical tomogram of the state $\ket{\psi_l}$ as
\begin{eqnarray}
\omega_{\alpha}\left(X_{\theta},\theta,t=\pi/l\chi\right)
=\frac{1}{\sqrt{\pi}} \modu{\sum_{r=0}^{l-1}\,f_{r,l} \exp{\left[-\frac{X_{\theta}^2}{2}-\frac{\modu{\alpha}^2}{2}-\frac{\alpha_{r,l}^2\,e^{-i2\theta}}{2}+\sqrt{2}\,\alpha_{r,l}\,X_{\theta}\,e^{-i\, \theta}\right]}}^2,
\label{optCS_l}
\end{eqnarray}
where
\begin{eqnarray}
\begin{aligned}[c]
f_{r,l}= \left\{ 
  \begin{array}{l l}
   f_{r}^{(o)} & \quad \text{if $l$ is odd}\\
   f_{r}^{(e)} & \quad \text{if $l$ is even}
  \end{array} \right.
\end{aligned}
\quad \text{and} \quad 
\begin{aligned}[c]
\alpha_{r,l}= \left\{ 
  \begin{array}{l l}
   \alpha\, e^{-i\,2 \pi r /l } & \quad \text{if $l$ is odd}\\
   \alpha\,e^{-i\,\pi (2 r -1)/l } & \quad \text{if $l$ is even}
  \end{array} \right..
\end{aligned}
\end{eqnarray}
Figures \ref{fig:optCS}(b), \ref{fig:optCS}(c) and \ref{fig:optCS}(d), show the optical tomograms of the state $\ket{\psi_l}$ for $l=2,\,3$, and $4$, corresponding to the $2,\,3$, and $4$-sub-packet fractional revivals of the initial coherent state, respectively.   The value of field strength $\modu{\alpha}^2$ used for plotting the tomograms in figures  is $20$. Figure~\ref{fig:optCS}(b) shows the optical tomogram of the state $\ket{\psi_2}$, which is  superposition of the coherent states $\ket{i\alpha}$ and $\ket{-i\alpha}$ with weights $(1-i)/2$ and $(1+i)/2$ (Fourier expansion coefficients in Eq.~(\ref{cs_l})), respectively. This optical tomogram of the state $\ket{\psi_2}$ is a   structure with two sinusoidal strands.
 Thus, a structure with two sinusoidal strands  in the optical tomogram of the time evolved state for an initial coherent state at $\trev/2$ is  a signature of $2$-sub-packet fractional revival. The quantum interference regions between the states $\ket{i\alpha}$ and $\ket{-i\alpha}$ are reflected in the optical tomogram of the state $\ket{\psi_2}$ at locations in the $X_{\theta}-\theta$ plane, where the two sinusoidal strands intersects, showing large oscillation in the optical tomogram.

The optical tomogram of the state $\ket{\psi_3}$, which is a state at $3$-sub-packet fractional revival,  plotted in Fig.~\ref{fig:optCS}(c)  displays a structure with three sinusoidal strands.  Similarly, the optical tomogram of the state  $\ket{\psi_4}$, which is a state at $4$-sub-packet fractional revival,  plotted in Fig.~\ref{fig:optCS}(d) shows a structure with four sinusoidal strands. We repeated the analysis for higher order fractional revivals ($l>4$) and found the general result that the optical tomogram of the time evolved state at $l$-sub-packet fractional revival time shows a structure with $l$ sinusoidal strands. 

During the evolution of coherent state $\ket{\alpha}$, the wave packet may also show collapse at specific instants of time $t=T_{rev}/s$, where $s$ is any  irrational number \cite{robi}.    At the instant of collapse  the state $\ket{\psi(t)}$ is not  a finite superposition of coherent states.  It has been shown that such collapsed states of the fields are of great importance because of its high nonclassical nature and can give large amount of entanglement when these states are split on a beam splitter with vacuum in the second input port \cite{rohith2}. To study the nature of optical tomogram during the collapse of wave packet, we plot the optical tomogram in Eq.~(\ref{optCS}) at collapse time $t=\trev/\sqrt{2}$.  The optical tomogram at this instant is shown in Fig.~\ref{fig:optCS}(e). The sinusoidal strands are not visible in the optical tomogram for the collapsed state. Which implies that the optical tomogram of a collapsed state is qualitatively different from that of the state at the instants of fractional revivals. Figure \ref{fig:optCS}(f) shows the revival of the initial state  at $t=\trev$.  
We can conclude that signatures of revivals and fractional revivals are captured in the optical tomogram of the time evolved states. Optical tomogram at the instants of $l$-sub-packet fractional revivals   shows $l$  sinusoidal strands for an initial coherent state, which is having one strand in its optical tomogram.

\subsection{Evolution of $m$-photon-added coherent state}
\begin{figure}
\centering
\includegraphics[height=4 cm, width=16 cm]{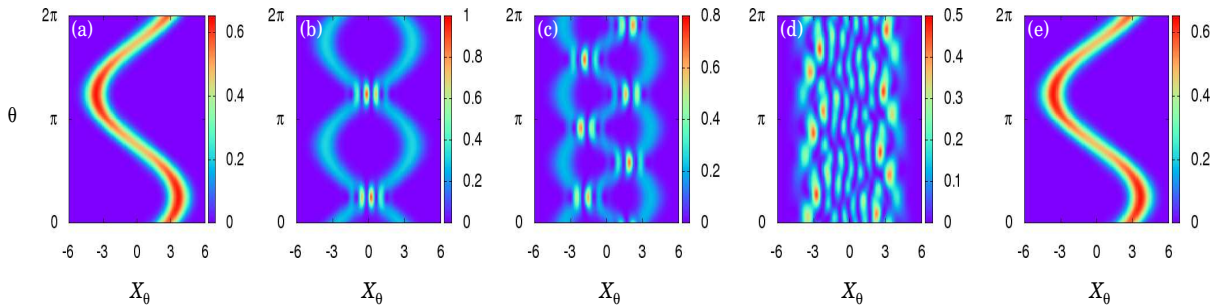}
\caption{Time evolved optical tomogram $\omega_{\alpha,1}\left(X_{\theta},\theta,t\right)$ of the  initial $1$-photon added coherent state at  (a) $t=0$, (b) $t=\trev/2$, (c) $t=\trev/3$, (d) $t=\trev/\sqrt{2}$, and (e) $t=\trev$, with field strength $\modu{\alpha}^2=5$. At $l$-sub-packet fractional revival time $t=\pi/l\chi$, the optical tomogram of the state shows a structure with $l$ sinusoidal strands. The structures with sinusoidal strands are completely absent in the optical tomogram for the collapsed state at time $t=\trev/\sqrt{2}$.}
\label{fig:optPACS}
\end{figure}

Here we consider the evolution of a nonclassical initial state, namely, an $m$-photon-added coherent state \cite {agarwal}
\begin{equation}
\ket{\alpha,m}=N_{\alpha,m}\, {a^{\dag}}^m \ket{\alpha},
\end{equation}
where $N_{\alpha,m}$ is the normalization constant and $m$ is the number of photons added to the coherent field $\ket{\alpha}$. One of the states of this family, namely, the $1$-photon-added coherent state has been experimentally produced by parametric down conversion process in a nonlinear crystal and the Wigner distribution of the state is reconstructed from the optical tomogram \cite{zavatta}. The Fock state expansion coefficient $C_n$ in Eq.~(\ref{initial}) for $m$-photon-added coherent state is
\begin{eqnarray}
C_n= \left\{ 
  \begin{array}{l l}
    0 & \quad \text{if $n<m$}\\
    \frac{e^{-\modu{\alpha}^2/2}\,\alpha^{n-m} \,\sqrt{n!}}{\sqrt{m!L_m(-\modu{\alpha}^2)}\,(n-m)!} & \quad \text{if $n\geq m$}
  \end{array} \right.,
\end{eqnarray}
where $L_m$ is the Laguerre polynomial of order $m$. The optical tomogram for $m$-photon-added coherent state have been theoretically investigated in \cite{korennoy}. The optical tomogram of $1$-photon-added coherent state given in Fig.~\ref{fig:optPACS}(a) displays a structure with single sinusoidal strand. It shows significant deviation of intensity along the sinusoidal strand, when it is compared with the optical tomogram of the coherent state given in Fig.~\ref{fig:optCS}(a). It is observed that the variation of the intensity becomes more pronounced as the value of $m$ increases. This is due to the increase in nonclassicality of the $m$-photon-added coherent state with increase in photon excitation number $m$ \cite{sudheesh3, usha}. The maximum intensity of the optical tomogram of $m$-photon-added coherent state along the $X_{\theta}$-axis occurs at $X_{\theta}=\sqrt{2 \aver{N}_m}\cos \delta$, where $\aver{N}_m$ is the mean photon number in the $m$-photon-added coherent state $\ket{\alpha,m}$. 

Time evolution of the initial $m$-photon-added coherent state under the Kerr Hamiltonian  shows revival and fractional revival at the same instants as in the case of initial coherent state \cite{sudheesh2}. Substituting $C_n$ in Eq.~(\ref{Genopticaltomo}), we get the time evolution of the optical tomogram for initial $m$-photon-added coherent state: 
\begin{eqnarray}
\omega_{\alpha,m}\left(X_{\theta},\theta,t\right)=\frac{e^{-\modu{\alpha}^2}}{m!L_m(-\modu{\alpha}^2)}\frac{e^{-X_{\theta}^2}}{\sqrt{\pi}}\modu{\sum_{n=m}^{\infty}\frac{\alpha^{n-m}\,e^{-i\chi t n(n-1)}}{(n-m)!\,2^{n/2}}e^{-in\theta} H_n\left(X_{\theta}\right)}^2.
\label{optPACS}
\end{eqnarray}
In Figs.~\ref{fig:optPACS}(b)-(e), we plot the optical tomograms $\omega_{\alpha,1}\left(X_{\theta},\theta,t\right)$ for the evolution of initial $1$-photon-added coherent state at different instants. Since the effect of photon addition to the coherent state $\ket{\alpha}$ is significant only for smaller field strengths, we choose $\modu{\alpha}^2=5$ for the plots. 
Fourier expansion of the time evolved state $\ket{\psi(t)}$ for an initial $m$-photon-added coherent state at   $l$-sub-packet fractional revival time $t=\pi/l\chi$, can be written as
\begin{eqnarray}
\begin{split}
\ket{\psi_l^{(m)}}= \left\{ 
  \begin{array}{l l}
    \sum_{r=0}^{l-1} f_{r}^{(o)}\, e^{-i\,2 \pi r m/l }\ket{\alpha\, e^{-i\,2 \pi r /l },m} & \text{if $l$ is odd}\\
    \sum_{r=0}^{l-1} f_{r}^{(e)}\, e^{-i\,\pi m(2 r -1)/l }\ket{\alpha\,e^{-i\,\pi (2 r -1)/l } ,m} & \text{if $l$ is even.}
  \end{array} \right.
  \end{split}
  \label{pacs_l}
\end{eqnarray}
Each of the ket vectors appearing in the right hand side of Eq.~(\ref{pacs_l}) is an $m$-photon-added coherent state, and the state $\ket{\psi_l^{(m)}}$ is a superposition of $l$ $m$-photon-added coherent state. The optical tomogram of the state $\ket{\psi_l^{(m)}}$  for an initial $1$-photon-added coherent state at  $2$ and $3$-sub-packet fractional revivals  are shown in Figs.~\ref{fig:optPACS}(b) and \ref{fig:optPACS}(c), respectively. The optical tomogram displays a structure with two sinusoidal strands  at $2$-sub-packet fractional revival, and is a structure with three sinusoidal strands  at $3$-sub-packet fractional revival.   The two sinusoidal strands in the optical tomogram  of the state $\ket{\psi_2^{(1)}}$ corresponds to the superposition of the  states $\ket{i\alpha,1}$ and $\ket{-i\alpha,1}$.  Similarly, the three sinusoidal strands in the optical tomogram of the state $\ket{\psi_3^{(1)}}$ corresponds to  the superposition of the states $\ket{\alpha,1}$, $\ket{\alpha e^{-i2\pi/3},1}$ and $\ket{\alpha e^{i2\pi/3},1}$. We repeated the analysis for higher values of photon excitation $m$ and for higher order fractional revivals ($l>3$) and found that at the instants of $l$-sub-packet fractional revivals,  the optical tomogram of the time evolved state for an  initial $m$-photon-added coherent state displays a structure with $l$ sinusoidal strands.  Figure \ref{fig:optPACS}(d) shows the optical tomogram of the time evolved state for an initial $1$-photon-added coherent state at collapse time $t=\trev/\sqrt{2}$. In this  case, optical tomogram does not show any sinusoidal strand. Again, like in the case of an initial coherent state,  the optical tomogram of the time evolved state at time $t=\trev$ shows the revival of the  initial $1$-photon-added coherent state. So far we have considered two types of initial states: coherent states and $m$-photon-added coherent states.   These states  are similar in the sense that both  are single wave packet with  a structure with single  sinusoidal strand in  their   optical tomograms. This is the reason for showing same number of sinusoidal strands at the instants of fractional revivals for these two kinds of initial  states. But they are completely different class of states because 
coherent states is a classical state and $m$-photon-added coherent state is nonclassical state. This difference is shown up in the intensity of sinusoidal strands in the optical tomogram. In the next section,  we consider   superposed   wave  packets, which are different from the initial states considered so far.

\subsection{Evolution of even and odd coherent states}
\begin{figure}
\centering
\includegraphics[height=8 cm,width=12cm]{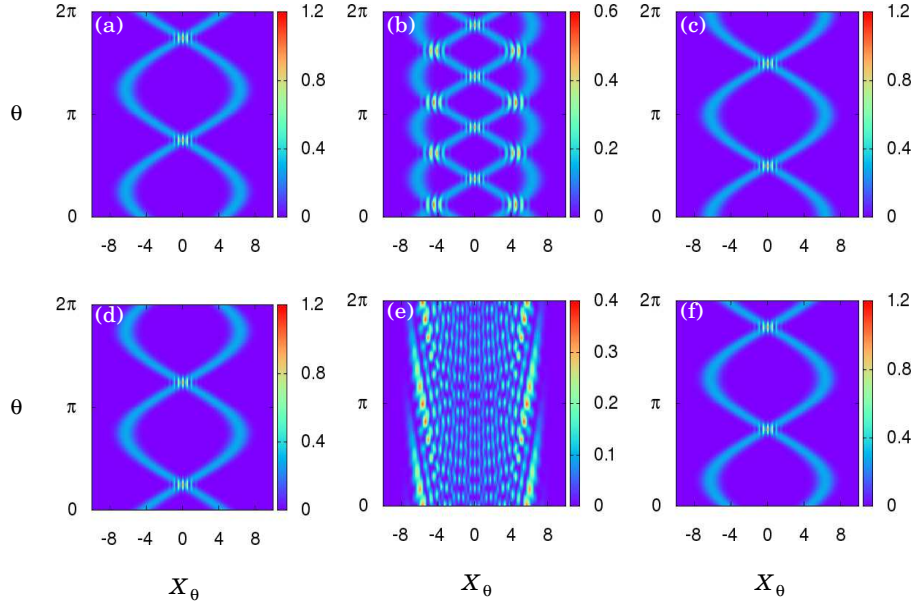}
\caption{Time evolved optical tomogram $\omega_{0}\left(X_{\theta},\theta\right)$  of the initial even coherent state with field strength $\modu{\alpha}^2=20$ at  (a) $t=0$, (b) $t=\trev/8$, (c) $t=\trev/4$, (d) $t=\trev/2$, (e) $t=\trev/\sqrt{2}$, and (f) $\trev$, respectively. At the instants of  $l$-sub-packet fractional revivals,  the optical tomogram of the time evolved state of the  initial even coherent state displays a structure with $2l$ sinusoidal strands.}
\label{fig:optEven}
\end{figure}

Consider the evolution of even and odd coherent states \cite{dodonov}, defined by
\begin{equation}
\ket{\psi(0)}_{h}=N_{h}\left[\ket{\alpha}+(-1)^h \ket{-\alpha}\right]\label{evenCS},
\end{equation}
where $h=0$ and  $1$, respectively. $N_{h}$ is an appropriate normalization constant. Fock state representation of even (odd) coherent state contains only the even (odd) photon excitations. The revival and fractional revivals during the evolution of initial even and odd coherent states in a Kerr media have been discussed in detail \cite{rohith2014}. It has been shown that \cite{rohith2014},  the time evolved state at $t=k\trev/4$, where $k=1,\,2$ and $3$, is a rotated initial wave packet. Also, the $l$-sub-packet fractional revival occurs at $t=j\trev/4l$ where $j = 1, 2,... , (4l - 1)$ for a given value of $l(> 1)$. Here $j$ and $4l$ are mutually prime integers. The Fock state expansion coefficient $C_n$ in Eq.~(\ref{initial}) for the even and odd coherent states is given by
\begin{eqnarray}
C_n= 2\,N_{h}\,e^{-\modu{\alpha}^2/2}\,\frac{\alpha^{n}}{\sqrt{n!}}\,\,\delta_{\left[\frac{n-h}{2}\right],\frac{n-h}{2}},
\end{eqnarray}
where $\delta$ is  Kronecker delta function, and $\left[x\right]$ is integer part of $x$. The symplectic tomography of even and odd coherent states have been discussed in \cite{mancini1996}. Inserting the value of the coefficient $C_n$ in Eq.~(\ref{Genopticaltomo}), we obtain the time evolved optical tomogram for initial even and odd coherent states:
\begin{eqnarray}
\omega_{h}\left(X_{\theta},\theta,t\right)=\frac{4\,N_{h}^2\,e^{-\modu{\alpha}^2}\,e^{-X_{\theta}^2}}{\sqrt{\pi}}\modu{\sum_{n=0}^{\infty}\frac{\alpha^{n}\,e^{-i\chi t\, n(n-1)}}{n!\,2^{n/2}}e^{-i n\theta} H_{n}\left(X_{\theta}\right)\,\delta_{\left[\frac{n-h}{2}\right],\frac{n-h}{2}}}^2.
\label{optEven}
\end{eqnarray}
 We focus on the evolution of an initial even coherent state $\ket{\psi(0)}_{0}$, but our analysis can be done for an initial odd coherent state as well. Figure~\ref{fig:optEven}(a) shows the optical tomogram of the  even coherent state with $|\alpha|^2=20$, which displays a structure with two sinusoidal strands. In Figs.~\ref{fig:optEven}(b)-(f), we plot the optical tomogram given in Eq.~(\ref{optEven}) at different instants during the evolution of the  initial even coherent state ($h=0$) in the medium. At the instants of rotated wave packets, the state is again a superposition of two coherent states. For example, at  $t=\trev/4$ and $\trev/2$, the Figs.~\ref{fig:optEven}(c) and \ref{fig:optEven}(d) shows the optical tomogram  of rotated wave packets. Optical tomogram shows a structure with two sinusoidal strands, as expected.  These tomograms  are qualitatively different from the the optical tomogram shown in Fig.~\ref{fig:optEven}(a). The locations of the sinusoidal strands, where the maximum intensity of the optical tomogram along $X_{\theta}$-axis occur, in these optical tomograms are shifted due to the phase space rotation of the quantum states during the evolution of initial even coherent state in the medium.  

Figure \ref{fig:optEven}(b) shows the optical tomogram of the time evolved state  at  $\trev/8$, which corresponds   to $2$-sub-packet fractional revival. It displays  a structure with four  sinusoidal strands, which is a signature of $2$-sub-packet fractional revival for the initial even coherent state (Note that the optical tomogram of the initial even coherent state itself is a structure with two sinusoidal strands). Time evolved optical tomogram for initial even and odd coherent states are also analysed at higher order fractional revival times  and we found that, at the instants of $l$-sub-packet fractional revivals, the optical tomogram of the time evolved state for the initial even and odd coherent state displays a structure with $2l$ sinusoidal strands. Figure \ref{fig:optEven}(e) shows the optical tomogram of a collapsed state at time $t=\trev/\sqrt{2}$,  which again confirm our result that  sinusoidal strands are absent in the optical tomogram of the collapsed state. The optical tomogram of the time evolved state at revival time is shown in Fig.~\ref{fig:optEven}(f).

\section{Effect of decoherence on optical tomogram}
\label{decoherence}
We have analyzed the optical tomograms of pure quantum states undergoing unitary evolution in the Kerr  medium so far. But the real optical nonlinearities are noisy and suffers various kinds of losses.  This leads to decoherence of the quantum states prepared in an experiment. Effect of decoherence on the time evolved states in a Kerr   media  has been theoretically studied using quasiprobability distributions \cite{milburn2,daniel}. In this section, we study the effect of environment induced decoherence on the optical tomogram of the time evolved state  $\ket{\psi(t)}$ given in Eq. (\ref{general_psi(t)}) at the instants of fractional revivals for the different initial states $\ket{\psi (0)}$ considered in the  earlier section.   Here we assume that the external environment consists of a collection of an infinite number of harmonic oscillators. Depending on the type of interaction between the single mode field and the environment, the decoherence of the quantum state can occur at least in two ways. First one is due to the photon absorption by the environment, also known as amplitude decay, and   the second one is due to the phase damping.  These two models are well-described by master equations. Consider the decoherence of the  state $\ket{\psi(t)}$ starts at $\tau=0$, and the  density matrix of the state at $\tau=0$ is given by  
\begin{equation}
\rho_t(\tau=0)=\ket{\psi(t)}\bra{\psi(t)}.
\end{equation}
The evolution of this state under decoherence can be represented in Fock basis as
\begin{equation}
\rho_t(\tau)=\sum_{n,n^\prime=0}^{\infty} {\rho_t}_{n,n^\prime}(\tau) \ket{n}\bra{n^\prime}, \label{decoh_densitymatrix}
\end{equation}
where the density matrix elements ${\rho_t}_{n,n^\prime}(\tau)$ can be calculated using  the appropriate master equations which describe
the amplitude decay and phase damping of the state \cite{gardiner}.
  The interaction with the external environment leaves the system in a mixed state, that is, the state given in Eq.~(\ref{decoh_densitymatrix}) is a mixed state for $\tau > 0$.

The optical tomogram of the state $\rho_t(\tau)$,   using Eq.~(\ref{opt_tomo_def}),  takes the form
\begin{equation}
\omega\left(X_\theta, \theta,t;\tau\right)=\sum_{n,n^\prime=0}^{\infty} {\rho_t}_{n,n^\prime}(\tau) \bra{X_\theta, \theta} n\rangle \langle n^\prime\ket{X_\theta, \theta}. \label{omega_decoh}
\end{equation}
The above expression for optical tomogram has been simplified to 
\begin{equation}
\omega\left(X_\theta, \theta,t;\tau\right)=\frac{e^{{-X_\theta}^2}}{\sqrt{\pi}}\sum_{n,n^\prime=0}^{\infty} {\rho_t}_{n,n^\prime}(\tau) \frac{H_n(X_\theta)\, H_{n^\prime}(X_\theta)}{2^{(n+n^\prime)/2}\,\sqrt{n!\,n^\prime!}} e^{-i\,(n-n^\prime)\theta}
\label{omega_Tau}
\end{equation}
where we have used the quadrature representation $\bra{X_\theta, \theta} n\rangle$ of the Fock state $\ket{n}$,
\begin{equation}
\bra{X_\theta, \theta} n\rangle=\frac{1}{\pi^{1/4}\,2^{n/2}}\frac{e^{{-X_\theta}^2/2}}{\sqrt{n!}} H_n(X_\theta) e^{-i\,n\,\theta}.\nonumber
\end{equation}
Evaluation of the optical tomogram $\omega\left(X_\theta, \theta,t;\tau\right)$ given in Eq. (\ref{omega_Tau}) involves the calculation of 
$ {\rho_t}_{n,n^\prime}(\tau)$, which depends  on the decoherence model. 
Next we  calculate $ {\rho_t}_{n,n^\prime}(\tau)$ for  amplitude decay model and   phase damping model. 

\subsection{Amplitude decay model}
\begin{figure}[h]
\centering
\includegraphics[width=12 cm]{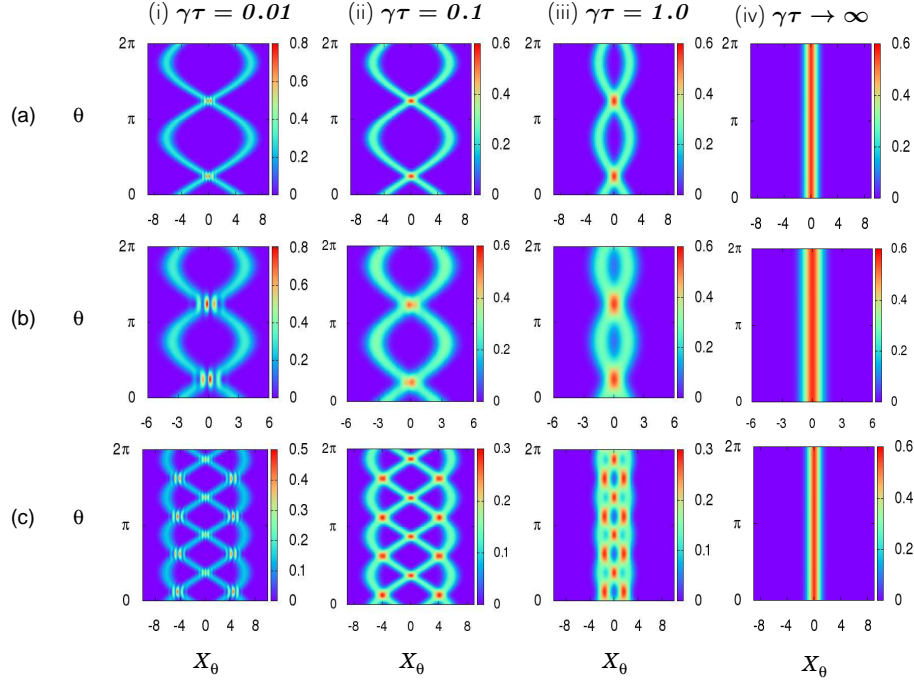}
\caption{Optical tomograms of the states at $2$-sub-packet fractional revival time  in the presence of amplitude damping  for initial (a) coherent state with $\modu{\alpha}^2=20$, (b) $1$-photon-added coherent state  with $\modu{\alpha}^2=5$, (c) even coherent state with $\modu{\alpha}^2=20$ at  (i) $\gamma \tau =0.01$, (ii) $\gamma \tau =0.1$, (iii) $\gamma \tau =1.0$ and (iv) $\gamma \tau\rightarrow\infty$.}
\label{fig:AmplitudeDamp}
\end{figure}

In this model, the interaction of the single mode field (mode $a$) with the environment modes $e_j$ under rotating wave approximation can be described by the Hamiltonian 
\bea
H_{amp}=\sum_j \hbar \gamma \left(a\, {e_j}^\dag+a^\dag e_j\right),
\eea
where $\gamma$ is the coupling strength of the mode $a$ with the environment. In the  Born-Markov approximation, the density matrix $\rho_t$ in the interaction picture   obeys the zero temperature master equation:
\begin{equation}
\frac{d \rho_t }{d \tau}=\gamma \left(2 a \rho_t a^\dag-a^\dag a \rho_t -\rho_t a^\dag a  \right). \label{master}
\end{equation}
The  matrix elements of $ {\rho_t}$ for an  arbitrary initial state  ${\rho_t}(\tau=0)$ is calculated in the Fock basis using  \cite{biswas}: 
\begin{equation}
{\rho_t}_{n,n^\prime}(\tau)=e^{-\gamma\tau(n+n^\prime)} \sum_{r=0}^{\infty} {\binom{n+r}{r}}^{1/2} {\binom{n^\prime+r}{r}}^{1/2} {\left(1-e^{-2\gamma \tau}\right)}^{r} {\rho_t}_{n,n^\prime}(\tau=0).\label{master_solution}
\end{equation}
It should be noted that all the density matrix elements ${\rho_t}_{n,n^\prime}(\tau)$ except those corresponds to $n=n^\prime=0$, decay exponentially to zero.  In the long time limit (i. e. $\tau\rightarrow\infty$) only the vacuum state will survive under amplitude decoherence, that is 
\bea
{\rho_t}(\tau\rightarrow\infty)=\ket{0}\bra{0}.
\label{rho_tau_infty}
\eea 
Using Eq.~(\ref{master_solution}), we calculate the   matrix elements ${\rho_t}_{n,n^\prime}(\tau)$  for the different initial states, $\ket{\psi(0)}$, considered earlier as follows:\\
(a) For the initial coherent state and $m$-photon-added coherent state,
\begin{eqnarray}
\begin{split}
{\rho_t}_{n,n^\prime}(\tau)=&\frac{e^{-\modu{\alpha}^2-\gamma\tau(n+n^\prime)}}{m!\,L_m\left(-\modu{\alpha}^2\right)}\sum_{r=0}^{\infty} {\binom{n+r}{r}}^{1/2} {\binom{n^\prime+r}{r}}^{1/2} {\left(1-e^{-2\gamma \tau}\right)}^{r}\\
 &\times\frac{\alpha^{n+r-m} \, {\alpha^\ast}^{n^\prime+r-m}\,\sqrt{(n+r)!\,(n^\prime+r)!}}{(n+r-m)!\,(n^\prime+r-m)!}
e^{-i\,\chi\,t\left[(n+r)(n+r-1)-(n^\prime+r)(n^\prime+r-1)\right]},
\end{split}
\label{decoh_density_elements_PACS}
\end{eqnarray}
($m=0$ corresponds to the initial coherent state).\\
(b) For the  initial even and odd coherent states,

%
   

\begin{eqnarray}
\label{decoh_density_elements_ECS}
{\rho_t}_{n,n^\prime}(\tau)&=4\,N_{h}^2\,e^{-\modu{\alpha}^2}e^{-\gamma\tau(n+n^\prime)}\sum_{r=0}^{\infty} {\binom{n+r}{r}}^{1/2} {\binom{n^\prime+r}{r}}^{1/2} {\left(1-e^{-2\gamma \tau}\right)}^{r}\,\frac{\alpha^{n+r} \, {\alpha^\ast}^{n^\prime+r}}{\sqrt{(n+r)!\,(n^\prime+r)!}}\nonumber\\
 &\times e^{-i\,\chi\,t\left[(n+r)(n+r-1)-(n^\prime+r)(n^\prime+r-1)\right]}\,\delta_{\left[\frac{n+r-h}{2}\right],\left(\frac{n+r-h}{2}\right)}\,\delta_{\left[\frac{n^\prime+r-h}{2}\right],\left(\frac{n^\prime+r-h}{2}\right)}.
\end{eqnarray} 
Optical tomograms of the initial $m$-photon-added coherent states and even and odd   coherent states can be obtained using 
the Eqs. (\ref{decoh_density_elements_PACS}) and (\ref{decoh_density_elements_ECS}),  respectively  in Eq.~(\ref{omega_Tau}).
In the following, we analyse   the amplitude damping of the state at $2$-sub-packet fractional revival time for different initial states. Figure ~(\ref{fig:AmplitudeDamp})  shows  the optical tomograms of the state   in the presence of amplitude damping at different  times $\gamma\tau$ (scaled time) for initial (a) coherent state ($m=0$), (b) $1$-photon-added coherent state, and (c) even coherent state. 
The structures with sinusoidal strands are not lost when  the interaction of the state with the environment is for a short duration of time  (for example, when   $\gamma\tau=0.1$).  
The sinusoidal strands come close together and get distorted  with increase in   time $\gamma\tau$,  and they merge together for large $\gamma\tau$. Figures  \ref{fig:AmplitudeDamp}(a) and (b) shows merging of two sinusoidal strands for initial coherent state and $1$-photon-added coherent state, respectively.   Plots  in  Fig. ~(\ref{fig:AmplitudeDamp})(c) shows merging of four sinusoidal strands for initial even coherent state.
 The merging of the sinusoidal strands with increase in  time $\gamma\tau$ is due to the decay of amplitude of the quantum state due to photon absorption by the environment. 
All the initial states considered above  decay to the vacuum state in the long time limit, i. e. when $\gamma \tau\rightarrow\infty$,  and  the corresponding optical tomogram is given by  
\bea
\omega\left(X_\theta,\theta,t;\tau\rightarrow\infty\right)=\frac{1}{\sqrt{\pi}}\,e^{-{X_\theta}^2}.
\label{Opt_vacuum}
\eea 
The above optical tomogram $\omega\left(X_\theta,\theta,t;\tau\rightarrow\infty\right)$  is a structure with single 
straight strand in $X_\theta-\theta$ plane and   the last column of the Fig. \ref{fig:AmplitudeDamp} confirms it. Another important fact is that,  the oscillations in the optical tomogram at interference regions of the sinusoidal strands decreases with increase in decoherence time $\gamma\tau$, which can be observed in the Fig.~\ref{fig:AmplitudeDamp}. 
We repeated the analysis described above for the states at  the instants of higher oder fractional revivals and found  similar results.

\subsection{Phase damping model}
In the phase damping model, the interaction between the system (represented by the mode $a$)  and the environment modes $e_j$ can be modelled by the Hamiltonian \cite{gardiner} 
\bea
H_{ph}=\sum_j \hbar \kappa \left(A {e_j}^\dag+A^\dag  e_j\right),
\eea
where $A=a^\dag a$ and $\kappa$ is the coupling constant. In this case, the interaction with environment causes no loss of energy of the system but the information about the relative phase of the energy eigenstates are lost. The Markovian dynamics of the state  $\rho_t$  is described by the zero temperature master equation
\begin{equation}
\frac{d \rho_t}{d \tau}=\kappa \left(2 A \rho_t A^\dag-A^\dag A \rho_t -\rho_t A^\dag A  \right). \label{master_phase}
\end{equation}
The matrix elements of $\rho_t$ for an arbitrary initial state ${\rho_t}(\tau=0)$ is given by \cite{gardiner}
\bea
{\rho_t}_{n,n^\prime}=\exp\left[-\kappa\left(n-n^\prime\right)^2\,\tau\right]{\rho_t}_{n,n^\prime}(\tau=0).
\label{master_ph_solution}
\eea
Here, the diagonal matrix elements  do not decay due to phase damping. Using Eq.~(\ref{master_ph_solution}) we calculate matrix elements of ${\rho_t}(\tau)$ in the presence of phase damping for different initial states as follows:\\
(a) For the initial coherent state and $m$-photon-added coherent states,
\begin{eqnarray}
\begin{split}
{\rho_t}_{n,n^\prime}(\tau)=\frac{e^{-\left(n-n^\prime\right)^2\kappa\tau}e^{-\modu{\alpha}^2}\alpha^{n-m} \, {\alpha^\ast}^{n^\prime-m}\,\sqrt{n!\,n^\prime!}}{m!\,L_m\left(-\modu{\alpha}^2\right)\,(n-m)!\,(n^\prime-m)!}e^{-i\,\chi\,t\left[n(n-1)-n^\prime(n^\prime-1)\right]}.
\end{split}
\label{PhDec_density_elements_PACS}
\end{eqnarray}
($m=0$ corresponds to  initial coherent state).\\
(b) For the  initial even and odd coherent states, 
\begin{eqnarray}
{\rho_t}_{n,n^\prime}(\tau)=\frac{4\,N_{h}^2
e^{-\left(n-n^\prime\right)^2\kappa\tau-\modu{\alpha}^2}\alpha^{n}
{\alpha^\ast}^{n^\prime}}{\sqrt{n!\,n^\prime!}}
e^{-i\,\chi
t\left[n(n-1)-n^\prime(n^\prime-1)\right]}\,\delta_{\left[\frac{n-h}{2}\right],\frac{n-h}{2}}\,\delta_{\left[\frac{n^\prime-h}{2}\right],\frac{n^\prime-h}{2}}.
\label{PhDec_density_elements_ECS}
\end{eqnarray}
Substituting Eqs.  (\ref{PhDec_density_elements_PACS}) and  (\ref{PhDec_density_elements_ECS}) in Eq.~(\ref{omega_Tau}) gives the optical tomogram of the time evolved state $\ket{\psi(t)}$ under phase damping for  initial $m$-photon-added coherent state, and even and odd coherent states, respectively. In the following, we consider  the phase damping of the state at $2$-sub-packet fractional revival time  for different initial states.  Figure \ref{fig:PhaseDamp} shows the optical tomograms of the state    in the presence of phase damping at different  times $\kappa\tau$ (scaled time) for initial (a) coherent state ($m=0$), (b) $1$-photon-added coherent state, and (c) even coherent state. 
In contrast to the amplitude damping, the phase damping shows a faster decay of the sinusoidal strands in the optical tomogram of the states. 
The sinusoidal strands in the optical tomogram of the state  retains its structure   only for a short  time $\kappa\tau$. The faster  decay of the state   is very noticeable in the case of an initial even coherent state, and  this aspect is displayed  in  Fig.  \ref{fig:PhaseDamp}(c).
 In the long time limit $\rho_t(\tau\rightarrow\infty)$, the optical tomograms  $\omega\left(X_\theta, \theta,t;\tau\rightarrow\infty\right)$ of different initial states are: \\
 (a) For initial coherent state and $m$-photon-added coherent states, 
 \begin{equation}
\omega\left(X_\theta, \theta,t;\tau\rightarrow\infty\right)=\frac{e^{{-X_\theta}^2-\modu{\alpha}^2}}{\sqrt{\pi}\,m!\,L_m\left(-\modu{\alpha}^2\right)}\sum_{n=m}^{\infty} \frac{\modu{\alpha}^{2(n-m)}\,H_{n}^{2}(X_\theta)}{2^{n}\,\left[(n-m)!\right]^2}
\end{equation}
 ($m=0$ corresponds to  initial coherent state).\\
 (b) For even and odd coherent states,

\begin{equation}
\omega\left(X_\theta, \theta,t;\tau\rightarrow\infty\right)=\frac{4\,N_{h}^{2}\,e^{{-X_\theta}^2-\modu{\alpha}^2}}{\sqrt{\pi}}\sum_{n=0}^{\infty} \frac{\modu{\alpha}^{2n}\,H_{n}^{2}(X_\theta)}{2^{n}\,\left(n!\right)^2}\,\delta_{\left[\frac{n-h}{2}\right],\frac{n-h}{2}}.
\end{equation}
Both the above tomograms are independent of the phase $\theta$, and it is displayed in  the last column of the  Fig.~\ref{fig:PhaseDamp}.
We have repeated the above analysis for higher order fractional revivals and found similar results.  Optical tomogram of the states, in the long time scales,  shows completely different structure for amplitude damping and phase damping models of the decoherence. This can be used to understand the type of interaction  the system is having with its environment. 
 \begin{figure}[h]
\centering
\includegraphics[width=12 cm]{fig5.eps}
\caption{Optical tomograms of the states at $2$-sub-packet fractional revival time in the presence of phase damping  for initial (a) coherent state with $\modu{\alpha}^2=20$, (b) $1$-photon-added coherent state with $\modu{\alpha}^2=5$, (c) even coherent state with $\modu{\alpha}^2=20$, at  (i) $\kappa \tau =0.01$, (ii) $\kappa \tau =0.1$, (iii) $\kappa \tau =0.3$ and (ii) $\kappa \tau \rightarrow \infty$.}
\label{fig:PhaseDamp}
\end{figure}
\section{Conclusion}
\label{conclusion}
We have studied the optical tomography of  different kinds of wave packets evolving in a Kerr medium. Exact analytical expression for the optical tomogram of the quantum states at any instant during the evolution of a generic initial wave packet is derived in terms of Hermite polynomials. Time evolution of the optical tomogram is analysed for three specific initial states:  a coherent state, an $m$-photon-added coherent state, and even and odd coherent states. We have shown that the signatures of revival and fractional revivals are captured in the optical tomograms of the quantum states. The optical tomogram of the time evolved state at the instants of fractional revivals shows structures with sinusoidal strands.  In general, the number of sinusoidal strands in the optical tomogram of the time evolved state at $l$-sub-packet fractional revivals are  $l$ times the number of sinusoidal strands present in the optical tomogram of the initial state considered. There are no sinusoidal strands present when the initial state is collapsed during the evolution. 
Interactions of the system with  its environment are inevitable in an real experimental settings, and  we found the manifestations of decoherence  directly  in the optical tomogram. 
The results obtained in this paper may be very useful for the experimental   characterisation of revivals and fractional revivals because of the following reasons: (1) We have shown the signatures of fractional revivals directly in the optical tomogram of the states. There is no need to reconstruct the density matrix or the quasiprobability distribution from the experimentally obtained optical tomogram to study fractional revivals, so that no need to concern about the  error that can accumulate during the  reconstruction process.  (2)  The analytical results obtained can be used  to verify and compare the optical tomogram generated in an homodyne measurement. (3) The theoretical results on decoherence can be used to find out how much the decoherence models really capture  the effects of environmental interactions in an actual experimental settings.


\begin{thebibliography}{60}
\bibitem{robi} R. W. Robinett,  Phys. Rep. {\bf 392}, 1 (2004).
\bibitem{aver} I. Sh. Averbukh and N. F. Perelman, Phys. Lett. A {\bf 139}, 449 (1989).

\bibitem{rempe} G. Rempe, H. Walther, and N. Klein, Phys. Rev. Lett. {\bf 58}, 353 (1987).
\bibitem{yeazell1990} J. A. Yeazell, M. Mallalieu, and C. R. Stroud, Jr., Phys. Rev. Lett. {\bf 64}, 2007 (1990).
\bibitem{yeazell} J. A. Yeazell and C. R. Stroud, Jr., Phys. Rev. A {\bf 43}, 5153 (1991).
\bibitem{Meacher} D. R. Meacher, P. E. Meyler, I. G. Hughes, and P. Ewart, J. Phys. B: At. Mol. Opt. Phys. {\bf 24}, L63 (1991).
\bibitem{vrakking} M. J. J. Vrakking, D. M. Villeneuve, and Albert Stolow, Phys. Rev. A {\bf 54}, R37 (1996).
\bibitem{greiner} M. Greiner, O. Mandel, T. W. Hansch, and I. Bloch, Nature {\bf 419}, 51 (2002).
\bibitem{Matsukevich} D. N. Matsukevich, T. Chaneli\'{e}re, S. D. Jenkins, S.-Y. Lan, T. A. B. Kennedy, and A. Kuzmich, Phys. Rev. Lett. {\bf 96}, 033601 (2006).




\bibitem{yurke} B. Yurke and D. Stoler, Phys. Rev. Lett. {\bf 57}, 13 (1986).
\bibitem{miranowicz} A. Miranowicz, R. Tana{\'s}, and S. Kielich, Quantum Opt. {\bf 2}, 253 (1990).
\bibitem{paprzycka}  M. Paprzycka and R. Tana{\'s}, Quantum Opt. {\bf 4}, 331 (1992).
\bibitem{tara} K. Tara, G. S. Agarwal, and S. Chaturvedi, Phys. Rev. A {\bf 47}, 5024 (1993).
\bibitem{cerf} N. J. Cerf, A. Ipe, and X. Rottenberg, Phys. Rev. Lett. {\bf 85}, 1754 (2000).
\bibitem{shapiro} E. A. Shapiro, M. Spanner, and M. Y. Ivanov, Phys. Rev. Lett. {\bf 91}, 237901 (2003).
\bibitem{kirchmair} G. Kirchmair, B. Vlastakis, Z. Leghtas, S. E. Nigg, H. Paik, E. Ginossar, M. Mirrahimi, L. Frunzio, S. M. Girvin, and R. J. Schoelkopf, Nature {\bf 495}, 205 (2013).
\bibitem{jex} I. Jex and  A. Or\l{}owski,  J. Mod. Opt. {\bf 41}, 2301 (1994). 
\bibitem{vaccaro} J. A. Vaccaro and A. Or\l{}owski, Phys. Rev. A {\bf 51}, 4172 (1995).
\bibitem{miranowicz2} A. Miranowicz, J. Bajer, M. R. B. Wahiddin and  N. Imoto, J. Phys. A. Math. Gen. {\bf 34}, 3887 (2001).
\bibitem{sudheesh2004} C. Sudheesh, S. Lakshmibala, and V. Balakrishnan, Phys. Lett. A {\bf 329}, 14 (2004).
\bibitem{romera2007} E. Romera and F. de los Santos, Phys. Rev. Lett. {\bf 99}, 263601 (2007).
\bibitem{romera2008} E. Romera and F. de los Santos, Phys. Rev. A {\bf 78}, 013837 (2008).
\bibitem{rohith2014} M. Rohith, and C. Sudheesh, J. Phys. B: At. Mol. Opt. Phys. {\bf 47}, 045504 (2014).
\bibitem{leonhardt} U. Leonhardt, {\it Measuring the quantum State of Light}, Cambridge University Press, cambridge (1997).
\bibitem{vogel} K. Vogel and H. Risken, Phys. Rev. A {\bf 40}, 2847 (1989).
\bibitem{mancini1995} S. Mancini, V. I. Man'ko, and P. Tombesi, Quantum Semiclass. Opt. {\bf 7}, 615 (1995).
\bibitem{ariano1996} G. M. D'Ariano, S. Mancini, V. I. Man'ko, and P. Tombesi, Quantum Semiclass. Opt. {\bf 8}, 1017 (1996).
\bibitem{omanko1997} O.  Man'ko and V. I. Man'ko, J. Russ. Laser Res. {\bf 18}, 407 (1997).
\bibitem{mancini1996} S. Mancini, V. I. Man'ko, and P. Tombesi, Phys. Lett. A {\bf 213}, 1 (1996).
\bibitem{manko1999} V. I. Man'ko and R. Vilela Mendes, Phys. Lett. A {\bf 263}, 53 (1999).
\bibitem{manko2005} V. I. Man'ko, G. Marmo, A. Simoni, A. Stern, and F. Ventriglia, Phys. Lett. A {\bf 343}, 251 (2005).
\bibitem{manko2006} V. I. Man'ko, G. Marmo, A. Simoni, A. Stern, E. C. G. Sudarshan, and F. Ventriglia, Phys. Lett. A {\bf 351}, 1 (2006).
\bibitem{Ibort} A. Ibort, V. I. Man'ko, G. Marmo, A. Simoni, and F. Ventriglia, Phys. Scr. {\bf 79}, 065013 (2009).
\bibitem{smithey} D. T. Smithey, M. Beck, M. G. Raymer, and A. Faridani, Phys. Rev. Lett. {\bf 70}, 1244 (1993).
\bibitem{lvovsky} A. I. Lvovsky and M. G. Reymer, Rev. Mod. Phys. {\bf 81}, 299 (2009).
\bibitem{bellini} M. Bellini, A. S. Coelho, S. N. Filippov, V. I. Man'ko, and A. Zavatta, Phys. Rev. A {\bf 85}, 052129 (2012).
\bibitem{milburn} G. J. Milburn, Phys. Rev. A {\bf 33},  674 (1986).
\bibitem{kita} M. Kitagawa and Y. Yamamoto, Phys. Rev. A {\bf 34}, 3974 (1986).
\bibitem{filippov} S. N. Filippov and V. I. Man'ko, Phys. Scr. {\bf 83}, 058101 (2011).
\bibitem{korennoy} Ya. A. Korennoy and V. I. Man'ko, Phys. Rev. A {\bf 83}, 053817 (2011).
\bibitem{adam} A. Miranowicz, M.  Paprzycka, A. Pathak, and F. Nori, Phys. Rev. A {\bf 89}, 033812 (2014).
\bibitem{rohith3} M. Rohith and C. Sudheesh, arXiv:1505.02698 [quant-ph], (2015).
\bibitem{barnett} S. M. Barnett and P. M. Radmore, {\it Methods in theoretical quantum optics}, Oxford University Press, Oxford, (1997).
\bibitem{rohith2} M. Rohith, R. Rajeev and C. Sudheesh, arXiv:1409.2643 [quant-ph], (2014) (to be published).
\bibitem{agarwal} G. S. Agarwal and K. Tara, Phys. Rev. A {\bf 43}, 492 (1991).
\bibitem{zavatta} A. Zavatta, S. Viciani, and M. Bellini, Science {\bf 306}, 660 (2004).
\bibitem{sudheesh3}C. Sudheesh, S. Lakshmibala, and  V. Balakrishnan, J. Opt. B: Quantum Semiclass. Opt. {\bf 7}, S728 (2005).
\bibitem{usha}  A. R. Usha Devi, R. Prabhu, and M. S. Uma,  Eur. Phys. J. D {\bf 40}, 133 (2006).  
\bibitem{sudheesh2} C. Sudheesh, S. Lakshmibala, and V. Balakrishnan, Europhys. Lett. {\bf 71}, 744 (2005). 
\bibitem{dodonov} V. V. Dodonov, I. A. Malkin, and V. I. Man'ko, Physica {\bf 72}, 597 (1974).
\bibitem{milburn2} G. J. Milburn and C. A. Holmes, Phys. Rev. Lett. {\bf 56},  2237 (1986).
\bibitem{daniel} D. J. Daniel and G. J. Milburn, Phys. Rev. A {\bf 39},  4628 (1989).
\bibitem{gardiner} C. W. Gardiner, {\it Quantum Noise}, Springer, Berlin, (1991).
\bibitem{biswas} A. Biswas and G. S. Agarwal, Phys. Rev. A {\bf 75}, 032104 (2007).

\end{thebibliography}
\end{document}